# Enhancing Decision Making Capacity in Tourism Domain Using Social Media Analytics

Supun Abeysinghe[#1], Isura Manchanayake[#2], Chamod Samarajeewa[#3], Prabod Rathnayaka[#4], Malaka J. Walpola[#5], Rashmika Nawaratne[*6], Tharindu Bandaragoda[*7], Damminda Alahakoon[*8]

[#]Department of Computer Science and Engineering, University of Moratuwa
Sri Lanka
[1]supun.14@cse.mrt.ac.lk
[2]isura.14@cse.mrt.ac.lk
[3]chamod.14@cse.mrt.ac.lk
[4]prabod.14@cse.mrt.ac.lk
[5]malaka@cse.mrt.ac.lk

[*]Research Centre for Data Analytics and Cognition, La Trobe University
Victoria, Australia
[6]B.Nawaratne@latrobe.edu.au
[7]t.bandaragoda@latrobe.edu.au
[8]d.alahakoon@latrobe.edu.au

*Abstract-Social media has gained an immense popularity over the last decade. People tend to express opinions about their daily encounters on social media freely. These daily encounters include the places they travelled, hotels or restaurants they have tried and aspects related to tourism in general. Since people usually express their true experiences on social media, the expressed opinions contain valuable information that can be used to generate business value and aid in decision-making processes. Due to the large volume of data, it is not a feasible task to manually go through each and every item and extract the information. Hence, we propose a social media analytics platform which has the capability to identify discussion pathways and aspects with their corresponding sentiment and deeper emotions using machine learning techniques and a visualization tool which shows the extracted insights in a comprehensible and concise manner. Identified topic pathways and aspects will give a decision maker some insight into what are the most discussed topics about the entity whereas associated sentiments and emotions will help to identify the feedback.*

*Index Terms-social media, tourism, insights, machine learning*

## I. INTRODUCTION

Nowadays social media plays a key role in lives of people in the world. It has disrupted the way people live their lives and it is widely adopted all over the world. As per the time of writing, Facebook which is the most commonly used social media website in the world has more than 2 billion monthly active users and Twitter has more than 334 million active users per month [1]. If the daily activity is considered, around 55 million status updates are made daily on average on Facebook [2] whereas more than 500 million tweets sent per day on Twitter [3]. Besides Facebook and Twitter, there is a number of other social media and community websites with similar statistics. With these huge numbers of daily activities, comes an immense pile of social media data which can be analyzed using big data analytics and machine learning to find interesting patterns.

Except for some other personal usages, social media is increasingly utilized by the public for expression of emotions and opinions on their day-to-day encounters. Especially tourists tend to share their experiences about certain attractions and other places in general through social media. Hence, these social data contain valuable information which can be extracted and then be used in decision-making processes. As an example, reviews of guests staying at a certain hotel can be used by its managers to improve the service provided. On a larger scale, the opinion of the people on a government decision can be used to take relevant measures to continue or alter the decision. Thus, it is obvious that these social data comprise of valuable information [4].

For a considerably smaller entity, associated data existing on social media might be a smaller quantity. In contrast, for a substantially larger entity, there would be an enormous quantity of data. Thus, manually inspecting each and every item and analyzing these data will be time-consuming and a cumbersome task. Moreover, it is not practical for a widely known entity. Hence, an automated approach for analyzing these data will be crucial. In designing such a system to generate insights from data, the main challenge lies in the fact that the insights are not in a structured format but encapsulated in free-text of social media posts. Thus, usage of trivial algorithms for information extraction will be not viable and machine learning techniques should have to be employed.

Though there are automated tools which can be used to analyze these types of data, their functionality is limited and a comprehensive generalized insight generation tool is still lacking. It is often said that one picture is worth more than a thousand words, in this case, the generated insights will be of no use if there is no way of conveying the idea to the end user in an effective manner. Thus, a tool which has the capability to automatically extract relevant data, generate insights by applying machine learning techniques to that data and visualizing those generated insights in an informative manner is proposed in this project. Our objective of this proposed tool is to lay the foundations for a comprehensive analytics platform which can be extended with other insight generations and visualizations in future.

In this study, we focus on the tourism domain given the proliferation of social media data available but the tool can be extended to be used in any other domain as well.

## II. LITERATURE REVIEW

Our work can be split into five subtasks; discussion pathway extraction, aspect extraction, aspect-based sentiment analysis, emotion detection, and visualizations. In the literature, there is no other tool which combines all these aspects into one single tool. Hence, the literature can be analyzed under each of those tasks.

### A. Discussion Pathway Extraction

Topic extraction or topic modeling is a largely researched area in Natural Language Processing (NLP). Nevertheless, topic modeling for short texts is comparatively more challenging compared to longer documents. This is mainly due to the sparsity of content in short texts. Most of the reviews and social media posts are relatively shorter texts, thus techniques for extracting topics from shorter texts are reviewed in this section. Since there is no sufficiently large annotated dataset for topics and manual labeling is expensive, unsupervised techniques are preferred.

Latent Dirichlet Allocation (LDA) [5] is one of the earliest approaches for topic modeling. It is a generative model which works under two underlying assumptions, which are documents are probability distributions over latent topics and topics are probability distributions over words. One limitation of this model is that it only gives better results when the document length is sufficiently large. An extension to this model is proposed in Author-Topic model [6] where author information is also incorporated into the generative process. The model aggregates short texts by the same author into a single large document and tries to identify topics. However, it highly relies on the assumption that all the set of short texts of an author is related to a single topic which is not true in the context of social media and reviews. In [7], the authors have identified a method to aggregate microblogs based on hashtags (in the case of tweets) and shown empirically that it yields better results compared to earlier approaches. Still, it does not cater for the possibility that a certain hashtag is used to discuss a diverse set of topics.

One other limitation of the above-identified approaches is that the topics or the number of topics are predefined. And the temporal aspect of topic altering and topic dependencies are not considered. Even though LDA based topic modeling that accompanies temporal aspects have been introduced in [8] and [9], all those methods lack detecting topic pathways and relationships among topics. In [10], authors have introduced a novel way of detecting topic pathways. They have used a refined version of the IKASL algorithm [11] which is an unsupervised algorithm that extracts topics in an incremental manner. This model has the capability to model topic pathways and relationships between topics in short texts.

### B. Aspect Extraction

Aspect extraction is another researched area in Natural Language Processing. It is on detecting the different aspects which have been spoken about in texts. This is directly related to aspect-based sentiment analysis and it is observed that many models for aspect-based sentiment analysis perform aspect extraction as well in an end to end manner.

Over the past few years, many different methods have been proposed for aspect extraction. Frequency-based methods [12] [13] are some of the oldest methods used. In this approach most frequent words in reviews usually, nouns and pronouns are considered to be aspects. In some methods [13], part-of-speech patterns are applied to filter the terms additional to the frequency terms as well. Another approach is using syntactic relations in words to determine the aspects [14]. As an example, the noun-adjective relationship between sentiment words and aspect words can be used to detect aspects after detecting sentiment words. Apart from the above mentioned frequency based methods, several researches have been carried out on using machine learning approaches for aspect detection. In [15], a linear chain Conditional Random Field (CRF) model is used to label the words of a sentence as aspects or not. Several unsupervised machine learning models [16] [17] using LDA [5] are used as well.

### C. Sentiment Analysis

Sentiment analysis is one of the most widely researched areas in Natural Language Processing. Over the years, many researchers have identified different approaches and algorithms to perform this task. Based on the task granularity sentiment analysis is subcategorized into three types [18]; document level, sentence level, and aspect level sentiment classification. Aspect level sentiment analysis considers sentiment expressed towards each aspect or target hence being the most granular.

Earlier approaches to sentiment analysis include rule based methods [19] and SVM based methods [20] [21] which usually requires handcrafted feature engineering. The performance of the above mentioned traditional machine learning models highly relies on the quality of handcrafted features. Usually, these handcrafted feature engineering requires domain expertise, hence can be expensive. Since the adaptation of Neural Networks (NNs) and Deep Learning to the field of NLP, the need of handcrafted features was declined [22]. These techniques emerged as powerful approaches to the task of sentiment analysis as well. There are few different approaches used in deep learning regime for the task of sentiment analysis. One approach is to train sentiment specific word embeddings [23] from neighboring text [24] [25] [26]. Another approach is to learn semantic composition of texts for modeling sentiment [27] [28] [29] [30]. This approach is not suitable for social media posts since most of them have a loose structure. One other approach is to perform sentence classification using Convolutional Neural Networks (CNNs) [31]. Long Short Term Memory Networks (LSTMs) [32] and Gated Recurrent Units (GRUs) which are variants of Recurrent Neural Networks (RNNs) have successfully used in this task as well [33] [34]. This approach combined with attention mechanisms [35] have achieved the state-of-the-art results in many benchmark datasets.

### D. Emotion Detection

Emotion detection in natural language texts has been a key area researched recently. The complexity of deriving the implicit expressions has opened many research opportunities. One approach is to use distant supervision techniques where Supun Abeysinghe[#1], Isura Manchanayake[#2], Chamod Samarajeewa[#3], Prabod Rathnayaka[#4], Malaka J. Walpola[#5], Rashmika Nawaratne[*6], Tharindu Bandaragoda[*7], Damminda Alahakoon[*8] emotion-related hashtags and emojis are used as annotations. Such models are trained and evaluated using noisy data since hashtags and emojis does not necessarily reflect the emotion expressed in sentences. Hence, reliability of such models is comparably lower. Another approach for emotion detection is to use transfer learning. The frequent usage of emojis can be seen in social media posts to express the emotion associated with the text. DeepMoji [36] has addressed emotion prediction using a proposed variant of transfer learning called chain-thaw where they have used a pre-trained model which predicts emoji occurrences in microblogs. The model is based on a bidirectional Long Short Term Memory (LSTM) Network [32] which was trained using a huge dataset of 1 billion tweets. Using the recently introduced [37] there have been supervised learning approaches based on deep learning for the task of emotion detection.

*E. Visualization*

Effective tracking and analysis of discussion pathways on social media are valuable in many different ways. But without proper visualization, it might not be very useful to a nontechnical person. Over the years researchers have come up with various different ways to visualize results and findings in an interactive and meaningful way.

[38] proposed a system to analyze opinion diffusion on social media in an interactive way. It is a composite visualization that combines a Sankey graph with an improved density map to visually convey the flow of opinions among users. The Sankey graph is used to visualize the flow of users among different topics in an event over time, which provides the necessary context for opinion diffusion analysis.

[39] proposed a similar system for analyzing various evolution patterns that emerge from multiple topics. They have developed TextFlow, an interactive visual analysis tool that helps users analyze how and why the correlated topics change over time. They have presented a topic mining model to track and connect topics over time. To produce a useful and visually pleasing layout, they have formulated it as a three-level directed acyclic graph and have solved it by using a force directed simulation. TextFlow also tightly integrates interactive visualization with topic modeling techniques to facilitate users to discover the evolution patterns at different levels of detail.

Both of the above systems provide visualization to show evolution of topic pathways over time, but both of them lack the ability to represent sentiments and deeper emotions associated with each discussion. Thus, a similar visualization tool is proposed where it can show deeper emotions associated with each discussion. For the aspect based sentiment visualizations a straightforward method is used.

### III. PROPOSED SYSTEM

We propose a social media analytics platform which has the capability to extract microblogs related to an entity or a place and then identify discussion pathways, topics and aspects occur in such microblogs and analyze sentiment and deeper emotions associated with each of those topics. As depicted in figure 1, our system mainly composed of three components. First, there is a data extraction and preprocessing component where the related data is extracted from publicly available sources (E.g. Twitter, TripAdvisor, Agoda etc.). Insight generation component comes next where artificial intelligence is used to generate useful insights. This component has four main parts; discussion pathway identification, aspect extraction, aspect-based sentiment analysis and emotion detection. Upon generating insights, the next component shows a visualization of such insights in a way that is easily comprehensible and interpretable to an end user. A comprehensive description of each component is given in the following subsections.

*A. Data Extraction*

As the system is mainly focused on the tourism domain, Trip Advisor, Booking.com and Twitter were selected as the main data sources mainly due to the large daily user activity and popularity. All these sources have a huge community involvement in reviewing and discussing tourism related topics. Twitter data was scraped using Twint [40] by specifying hash-tags (e.g. *#sigiriya, #lka*) and the locations using geographic coordinates and radius. A custom scraper build by us was used to extract data from TripAdvisor and Booking.com reviews.

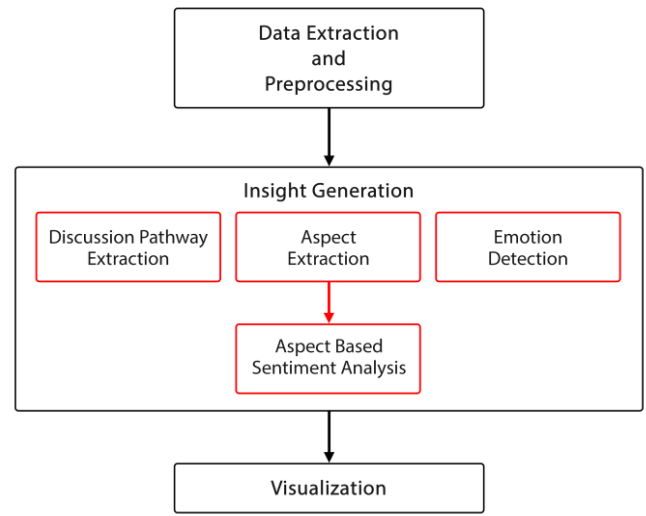

Fig. 1 Overview of the proposed system

*B. Preprocessing*

Preprocessing is a key step in the task pipeline. Typically, raw data extracted from social media posts and microblogs in general, have unstructured formats and informal language usages such as short terms, emojis, and hashtags. In order to clean the collected data, several preprocessing steps commonly used in Natural Language Processing (NLP) such as tokenizing, word normalization, spell correction, replacing emojis with corresponding words and word segmentation for hashtags and stop word removal were used. Preprocessing steps taken for each part in the insight generation component is different and have mentioned under each corresponding subsection.

1) Tweet Tokenizing

Tokenizing is the first and the most important step of preprocessing. Ability to correctly tokenize a tweet directly impacts the quality of a model. Since there is a large variety of vocabulary and expressions present in Twitter, it is a challenging task to correctly tokenize a given tweet. Twitter markup, emoticons, emojis, dates, times, currencies, acronyms, censored words (e.g. s**t), words with emphasis (e.g. *very*) are recognized during tokenizing and treated as separate tokens.

*2) Word Normalization*

After tokenizing, set of transformations are applied to the tokens. converting to lowercase, transforming URLs, usernames, emails, phone numbers, dates, times, hashtags to a predefined set of Tags. (e.g. @user1 → <user>). This helps to reduce the vocabulary size and generalize the tweet. We have used pretrained Glove word vectors for Twitter [41] which uses the same approach.

*3) Spell Correcting and Word Normalization*

As the last step in preprocessing, spell correcting and word segmentation were applied to hashtags. (e.g. #makeitrain → make it rain) By using the aforementioned procedure we can do a proper tokenization and generalization, to a given microblog.

*C. Discussion Pathway Identification*

Discussion pathway identification is the task of identifying topics discussed and how they evolve into different topics over time. Figure 2 shows an example discussion pathway evolution related to Sri Lanka. In the given scenario, people have talked about the riots happened in Kandy. After some time, that discussion topic evolved into two related topics where a set of people have talked about social media ban imposed by the government and another set of people about the curfew imposed in certain parts of the country. Meanwhile, a separate set of people were talking about the bad weather and subsequently about train services being stopped.

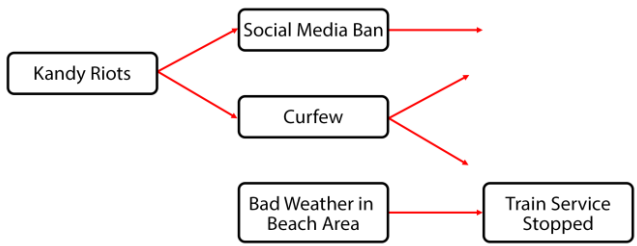

Fig. 2 An example for discussion pathway evolution over time

For the task of identifying discussion pathways, a refined version of Incremental Knowledge Acquisition and Self Learning (IKASL) [11] algorithm proposed in [10] is used. IKASL algorithm has the ability to cluster microblogs into groups based on semantic similarity. In contrast to other topic modeling algorithms, this algorithm does not require a predefined set of topics. Moreover, the algorithm has the ability to model dependencies between clusters in an incremental manner, allowing us to capture discussion pathway evolution. IKASL algorithm is an unsupervised learning algorithm and hence an annotated dataset is not required. However, several trials have been conducted using sample data for the purpose of tuning the model hyperparameters. At the preprocessing stage, stop words are removed since those words do not reflect the topic discussed in the text.

*D. Aspect Extraction*

Customer reviews have been a major feedback collecting source for many hotels and places. However, people tend to express their opinions on different aspects of a place or a hotel in a single review. Therefore, identifying each aspect on which the people have expressed their opinions is important. In our platform, we propose a model to extract the different aspects expressed in text reviews.

*E.g. - The food is very average...the Thai fusion stuff is a bit too sweet.*

The terms 'food' and 'Thai fusion stuff' will be detected as aspects of the above review. The detected aspects along with the text review are fed to the sentiment detection model to find the sentiments of the user on each aspect.

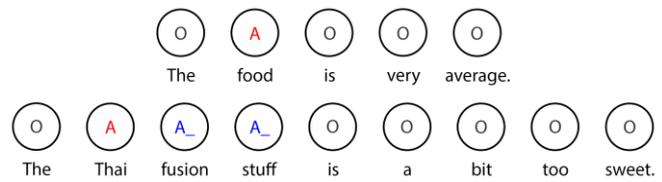

Fig. 3 Sentence tokens with their respective tags used for the aspect extraction task

Here, we have used a dataset presented in [42] of hotel reviews which contain a set of 1315 sentences with labeled aspect terms in English. Above example is an extracted review from the dataset. We have mapped the aspect extraction task into a sequence labeling problem by tagging each word using **O**, **A** and **A_** where **O** refers to non-aspect word, **A** refers to starting aspect word and **A_** refers to a intermediate word in a multi-word aspect term. Figure 3 shows the tags for the above example. Then we have trained an LSTM (Long Term Short Memory) [32] based sequence labeling model to identify aspect terms.

*E. Aspect based sentiment analysis*

Sentiment analysis has been a key area of research in the field of Natural Language Processing. The main goal of sentiment analysis is to identify the overall positive, negative or neutral opinion of authors in texts. However, when it comes to reviews, people tend to express different opinions or sentiments on different aspects in a single review. The example below expresses the opposing sentiments on different aspects.

*E.g. - The place is small and cramped but the food is fantastic.*

The above review contains two aspects 'place' and 'food'. The author has a negative sentiment towards the place while having positive sentiment towards the food. In such instances identifying a single sentiment for the whole review will result in inaccurate insights. Hence, we have used an aspect based sentiment analysis model proposed by [43]. It uses a segmentation attention based LSTM model to capture structural dependencies between the aspect terms and the sentiment expressions with a linear-chain conditional random field (CRF) layer.

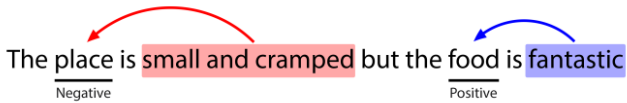

Fig. 4 Different sentiments expressed towards different aspects in a single sentence

*F. Emotion Detection*

The free speech has resulted in expressing true emotions in social media. Though sentiment analysis yields a high level view of the opinions of people, emotion analysis gives more fine-grained results in this task. Therefore, understanding emotions in social media is very useful as a feedback for any commercial service provider to make valuable decisions in tweaking their performance. We have used an attention based [44] approach for the task of emotion detection in social media. Recently released dataset by [37] was used for training and evaluation of our model. The model is based on a Gated Recurrent Neural Network (GRU) [45] which is a type of Recurrent Neural Network (RNN). The model uses Glove word embeddings [41] to accompany external knowledge about words and uses attention mechanisms. Ekphrasis tool introduced by [46] was used for tweet preprocessing for emotion detection. Tweet tokenizing, word normalization, spell correcting, replacing emoji with corresponding words and word segmentation for hashtags were done as preprocessing steps.

*G. Visualization*

The discussion pathways are visualized in a multiple parent tree structure so that one discussion can have multiple derived discussion topics as in the figure 5. The nodes in the tree represent a set of related posts for a specific discussion which are spread along a timeline. The color of nodes represents the prominent sentiment and emotion related to the discussion at the point of time.

The aspects for a specific hotel or place will be visualized as in the figure 6. The overall sentiment of different aspects of a place/hotel will be presented as a percentage of the number of posts which contain positive sentiment towards relevant aspects out of all the posts which contain those aspects. Furthermore, most frequent emotions associated with the place/hotel is displayed at the top.

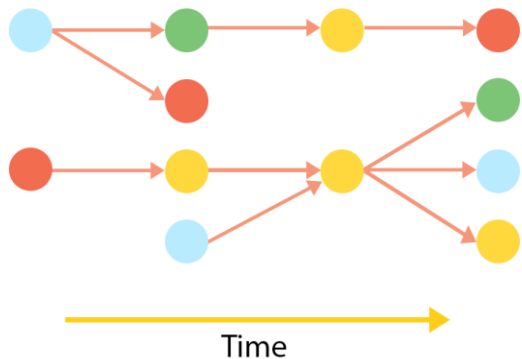

Fig. 5 Visualization of discussion pathways. Color in each node corresponds to the sentiment towards that topic.

IV. EXPERIMENTAL RESULTS

We have conducted some experiments to test the accuracies of the models used for aspect extraction, aspect-

Supun Abeysinghe[#1], Isura Manchanayake[#2], Chamod Samarajeewa[#3], Prabod Rathnayaka[#4], Malaka J. Walpola[#5], Rashmika Nawaratne[*6], Tharindu Bandaragoda[*7], Damminda Alahakoon[*8] based sentiment analysis and emotion detection. For evaluations, we have used existing annotated datasets.

For the task of aspect extraction, we have used the dataset presented in SemEval 2016 Task 5 [42] and the Table 1 shows the accuracy measures for the experiment. Based on the results mentioned in [42], our model surpasses the baseline models. However, our model does not achieve the state-of-the-art results, nevertheless improving the performance of aspect extraction is left as future work.

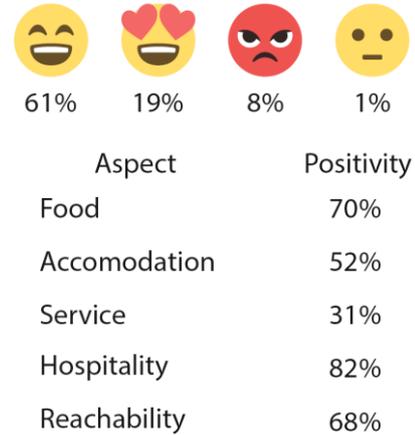

Fig. 6 Visualization of overall emotion towards an entity, aspects discussed related to that entity with their sentiment.

| Precision | Recall | F-score |
|---|---|---|
| 0.6162 | 0.6138 | 0.615 |

Table 1 Precision, recall and f-score for the model used for aspect extraction.

Moreover, we have done some experiments using some sample test cases as well. Table 2 shows that our model has the ability to identify multi-word aspects and multiple aspects residing inside single sentences as well.

| Review Sentence | Identified Aspects |
|---|---|
| The food was good, the place was clean and affordable. | food<br>place |
| I love margherita pizza | margherita pizza |
| Highly impressed from the quality of the food and the hospitality to the great night I had! | food<br>hospitality |
| Top quality food, but the service could have been much more better | food<br>service |

Table 2 Extracted aspects from example reviews.

Once the aspects are extracted, aspect-based sentiment analysis is done to identify sentiments towards each aspect. We have used the model proposed in [43] for this task. This model achieves the state-of-the-art results in the task of identifying sentiment towards an aspect. A comprehensive analysis about pros and cons of this model can be found in [43].

For the task of emotion detection, we have used the recently released dataset presented in [37] for SemEval 2018 Task 1. We have implemented a Bidirectional Gated Recurrent Neural Network (Bi-GRU) based model which uses attention mechanisms [44] for this task. Table 3 shows the results and we managed to surpass the state-of-the-art results using the proposed model. A comprehensive use case driven analysis of the platform and each component is left as future work.

| Precision | Recall | F-score |
|---|---|---|
| 0.67 | 0.71 | 0.68 |

Table 3 Precision, recall and f-score for the model used for emotion detection.

## V. Conclusion and Future Work

In this work, we have proposed a social media analytics platform for tourism domain which can be used to generate business value and aid in the decision-making process. The proposed system has the capability to identify discussion pathways and aspects discussed in microblogs and to analyze associated sentiments and emotions. For the current version of the system, only text data are collected and taken into the insight generation process. In this work, we have done a basic analysis of the proposed system and comprehensive analysis is identified as future work. Another potential future work can be an extension to the proposed system to accommodate information from images using computer vision techniques and using other information such as location data to generate more precise insights. Moreover, the proposed system can be extended into a comprehensive tourist social media analytics platform by including insights such as tourist profiling, tourism forecasting, tourist travel patterns, and geographical information analysis. Another potential future work is to employ the system into a different domain or a use case. This can be political domain and other business domains in general.

Supun Abeysinghe[#1], Isura Manchanayake[#2], Chamod Samarajeewa[#3], Prabod Rathnayaka[#4], Malaka J. Walpola[#5], Rashmika Nawaratne[*6], Tharindu Bandaragoda[*7], Damminda Alahakoon[*8]